\documentclass[letter]{aa} 
\usepackage{graphicx}
\usepackage{txfonts}
          
\begin{document}  
  
\title{On   the   origin  of   white   dwarfs  with   carbon-dominated
       atmospheres: the case of H1504+65}
  
\author{L. G. Althaus\inst{1,2,3},
        A. H. C\'orsico\inst{1,2,3},  
        S. Torres\inst{4,5}, \and 
        E. Garc\'{\i}a--Berro\inst{4,5}}   
  
\offprints{L. G. Althaus}  
  
\institute{Facultad de Ciencias Astron\'omicas y Geof\'{\i}sicas,  
           Universidad  Nacional de La Plata,
           Paseo del  Bosque S/N,  
           (1900) La Plata, 
           Argentina\
           \and
           Instituto de Astrof\'{\i}sica La Plata, 
           IALP, CONICET-UNLP,
           Argentina\
           \and 
           Member of the Carrera del Investigador Cient\'{\i}fico y 
           Tecnol\'ogico, CONICET, 
           Argentina\
           \and
           Departament de F\'\i sica Aplicada, 
           Escola Polit\`ecnica Superior de Castelldefels,
           Universitat Polit\`ecnica de Catalunya,  
           Av. del Canal Ol\'\i mpic, s/n,  
           08860 Castelldefels,  
           Spain\
           \and
           Institute for Space Studies of Catalonia,
           c/Gran Capit\`a 2--4, Edif. Nexus 104,   
           08034  Barcelona,  Spain\\  
\email{althaus,acorsico@fcaglp.unlp.edu.ar; santi,garcia@fa.upc.edu} }
    
\date{\today}  
  
\abstract{}
         {We explore  different evolutionary scenarios  to explain the
         helium  deficiency observed in  H1504+65, the  most  massive 
         known PG1159 star.}
         {We  concentrate mainly  on  the possibility  that this  star
         could be the result  of mass loss shortly
         after  the  born-again  and   during  the
         subsequent evolution through  the [WCL] stage. This possibility
         is sustained by recent observational evidence of extensive
         mass-loss events in Sakurai's object and is in line with the
         recent  finding that  such  mass losses  give  rise to  PG1159
         models with thin helium-rich  envelopes and large rates of period
         change, as  demanded by  the pulsating star  PG1159$-$035. We
         compute the post born again evolution of massive sequences by
         taking into account different mass-loss rate histories.}
         {Our   results  show   that  stationary   winds   during  the
         post-born-again  evolution  fail  to  remove  completely  the
         helium-rich envelope so as  to explain the helium deficiency
         observed in H1504+65. Stationary winds during the Sakurai and
         [WCL] stages only remove at  most half of the envelope surviving
         the violent hydrogen burning during the born-again phase.}
         {In view of our  results, the recently suggested evolutionary
         connection    born-again    stars   $\rightarrow$    H1504+65
         $\rightarrow$  white dwarfs  with carbon-rich  atmospheres is
         difficult to sustain unless  the whole helium-rich envelope could
         be ejected by non-stationary mass-loss episodes  
         during the Sakurai stage.}

  
\authorrunning{L. G. Althaus et al.}  
  
\titlerunning{On the origin of H1504+65}  
  
\maketitle  
  
   
\section{Introduction}  
\label{intro}  

Hydrogen (H)-deficient  PG1159 stars  are a  transition stage  between the
post-asymptotic   giant   branch  (AGB)   stars   and   most  of   the
H-deficient white  dwarfs.  Spectroscopic analyses  reveal that
PG1159  stars  are  characterized   by  a  peculiar  surface  chemical
composition,  with  most  of  them  exhibiting  helium (He)-,  carbon-  and
oxygen-rich  abundances  at  their  surfaces.  Inferred  surface  mass
abundances  are typically  $X_{\rm He}\simeq  0.33$,  $X_{\rm C}\simeq
0.5$ and  $X_{\rm O}\simeq 0.17$, though notable  variations are found
from star to  star (Dreizler \& Heber 1998;  Werner 2001).  Currently,
about 40 stars are members of the PG1159 spectroscopic family (Werner
\& Herwig  2006), which covers a  wide range of  surface gravities ---
$5.5 \la \log  g \la 8$ (in cgs units)  --- and effective temperatures
--- 75,000  K $\la  T_{\rm  eff}  \la$ 200,000  K.   PG1159 stars  are
thought to be  formed as a result of a born-again  episode, that is, a
very late thermal pulse (VLTP) experienced by a hot white dwarf during
its early cooling  phase --- see Sch\"onberner (1979)  and Iben et al.
(1983) for early  references --- or  a late thermal pulse  that occurs
during the  post-AGB evolution when  H burning is still  active (Bl{\"
o}cker 2001).

The variety of surface compositions  observed in most PG1159 stars has
been   successfully   explained  by   detailed   simulations  of   the
evolutionary  stages prior  to  their formation,  particularly of  the
born-again episode (Werner \& Herwig 2006; Rauch 2006).  For instance,
the high oxygen  abundance detected in the atmospheres  of these stars
and the presence  of neon lines in the optical  spectra of some PG1159
stars is  in line with  the improved evolutionary calculations  of the
born-again scenario  that incorporate convective  overshoot (Herwig et
al. 1999).  However, some  remaining cases still represent a challenge
to the stellar evolution theory, which cannot adequately explain their
origin. This is true for  the enigmatic star H1504+65, whose effective
temperature of $T_{\rm eff}=$200,000 K and large mass --- between 0.74
and $0.97\, M_{\sun}$ (Werner \& Herwig 2006; Werner et al.  2008) ---
make  it  the  hottest  and   most  massive  known  PG1159  star.   In
particular,  H1504+65 is a  H- and  He-deficient star,  and its
atmosphere is  dominated by carbon  and oxygen plus traces  of heavier
elements ---  $X_{\rm C}\simeq 0.48$, $X_{\rm  O}\simeq 0.48$, $X_{\rm
Ne}\simeq 0.02$,  $X_{\rm Mg}\simeq 0.02$  (Werner et al.   2004).  It
seems  like if we  were seeing  the naked  carbon-oxygen core  of this
star.  These unique characteristics  among the known PG1159 stars make
the evolutionary history of H1504+65 very intriguing.

The recent discovery of a cool white dwarf population with carbon-rich
atmospheres (Dufour et. al.  2007,  2008) has renewed the interest for
H1504+65 since  it opens  the possibility that  H1504+65 could  be the
progenitor of  this new white  dwarf population. However, there  is no
yet  a  clear consensus  about  the  evolutionary  history that  could
explain the origin of its He deficiency. In this work we explore 
  different evolutionary scenarios to explain  the origin of
H1504+65.  Most  importantly, we  assess the viability  that   mass 
loss during the evolutionary stages following the VLTP when the
star becomes a red gaint (Sakurai's stage)  and the following [WCL] 
stage could
be responsible for  the He deficiency observed in  H1504+65. This
is an  attractive idea  in view  of the recent  finding by  Althaus et
al.  (2008) that  PG 1159  stars  could be  characterized by  He--rich
envelopes markedly smaller than those predicted by the standard theory
of  stellar evolution  for the  formation of  PG1159 stars  (Herwig et
al. 1999; Miller Bertolami \& Althaus 2006).  
 
\section{Proposed evolutionary scenarios}

It  has  been speculated  that  H1504+65  could  have an  evolutionary
history completely  different from that of the  standard PG1159 stars.
In  particular, it  has  been  suggested ---  and  discarded ---  that
H1504+65 could  be a normal  H-rich post-AGB star  where strong
 mass losses  would  remove the He-rich intershell (Sch\"onberner
\& Bl\"ocker 1992). For this to be possible, very high mass-loss rates
--- of  about $10^{-7}\, M_{\sun}/{\rm  yr}$ ---  should occur  at low
luminosities.  However,  it is dubious that such  mass-loss rate could
be maintained by real stars during  the white dwarf stage. It has also
been suggested  (Werner 2001) the  possibility that H1504+65  could be
the result  of the evolution of a  heavy-weight intermediate-mass star
with  an initial  mass between  9 and  $11\, M_{\sun}$  that has
burned carbon  into oxygen  and neon (Ritossa  et al.   1996; Garc\'\i
a--Berro et al. 1997).  In this  scenario, He would be burned during a
{\sl late} carbon shell flash in the former Super-AGB phase.  However,
detailed  simulations  do not  seem  to  support  this.  Instead,  the
theoretical calculations  predict that carbon flashes  stop before the
thermally-pulsing  AGB phase  is reached (Ritossa  et al.   1996).  In
this scenario, the very massive envelope of these stars should be lost
before reaching the thermally  pulsing AGB phase, something that would
require  unrealistic mass-loss rates  during the  early AGB  phase. In
addition, the  resulting oxygen-neon core would be  too massive ($\sim
1.1\, M_{\sun}$) to  be compatible with the mass  of H1504+65 and with
the  spectral  features  observed   in  the  carbon-rich  white  dwarf
population detected by Dufour et al. (2007,2008). Another possibility would
be  that H1504+65 is  the product  of the  coalescence of  two typical
carbon-oxygen white dwarfs (Guerrero  et al. 2004). However, given the
short age of  H1504+65, this possibility seems very  unlikely. If this
were the case, this star should be rapidly spinning and, moreover, the
debris of the  disrupted secondary should form a  thick massive keplerian disk
around H1504+65, which is not observed.

A different  possibility is that  H1504+65 has an  evolutionary origin
not very different from those of the majority of PG1159 stars. Namely,
that its  progenitor has experienced  a born-again episode.   In fact,
Werner et al. (2004) have  raised the possibility that, because of the
large mass of this star, He has been completely burned into carbon and
oxygen during  the late He  shell flash.  Detailed simulations  of the
born-again  stage for  massive remnants  (Miller Bertolami  \& Althaus
2006) show,  however, that this  scenario is  not possible,  since our
calculations show  that even for the most  massive born-again sequence
($0.87\,  M_{\sun}$), most  of the  He content  before  the born-again
episode remains  unburnt.  This is a somewhat  expected result because
He  burning during  a late  He flash  is not  very different  from the
well-studied  cases  of thermal  pulses  on  the  AGB, for  which  the
available calculations do not show in any case a complete depletion of
the intershell He during the thermal pulse.

An alternative possibility, which we  explore in detail below, is that
H1504+65 is the result of  mass loss during the Sakurai's and
[WCL] stages.  The  possibility that a
large  fraction  of  the  He  envelope could  be   during  the
 Sakurai's stage is  attractive and  is supported  by two  recent and
different  pieces  of evidence.   The  first  one  is that  V4334  Sgr
(Sakurai's object) is displaying very strong mass-loss events 
(van Hoof et al.  2007).
The  second  piece  of  evidence  is the  recent  theoretical  finding
(Althaus et  al.  2008) that, as  a consequence of  such reported high
mass-loss  rates  during the  Sakurai  stage,  PG1159  stars could  be
characterized by much thinner He envelopes than traditionally thought.
Indeed, as shown in Althaus et al. (2008) a thin He--rich  envelope 
appears  to  be needed  to solve  the
longstanding discrepancy between the observed (see Costa \& Kepler
2008 for recent measurements) and theoretical rates of
period change of the  pulsating star PG1159$-$035.

\begin{table}
\caption{Main characteristics of the post-born-again remnants}
\centering
\begin{tabular}{@{}ccc}
\hline
\hline
$M/M_{\sun}$ & $M_{\rm He}/M_{\sun}$ & $\log(L_{\rm max}/L_{\sun})$\\
\hline
0.561 & 0.040  & 3.94 \\
0.564 & 0.037  & 3.95 \\
0.664 & 0.021  & 4.17 \\
0.741 & 0.014  & 4.32 \\
0.870 & 0.0058 & 4.58 \\ 
\hline
\hline
\end{tabular}
\end{table}

\section{Post-born-again evolutionary sequences}

The evolutionary calculations  for this work have been  done using the
{\tt LPCODE} evolutionary code  (Althaus et al.  2005), which computes
the  formation and  evolution  of white  dwarfs  through late  thermal
pulses  on  the  basis  of  a detailed  description  of  the  physical
processes relevant  for the calculation of  the violent proton-burning
phase during the born-again stage, particularly diffusive overshooting
and non-instantaneous mixing.   The code has also been  used by Miller
Bertolami \&  Althaus (2006)  who have computed  a full set  of PG1159
evolutionary  sequences  from  ZAMS  progenitors with  initial  masses
ranging  from 1.0 to  $5.5 \,  M_{\sun}$ to  the born-again  stage. In
particular, the $0.87 \, M_{\sun}$ model sequence, the most massive of
the  computed sequences, is  the result  of the  full evolution  of an
initially $5.5 \, M_{\sun}$ main sequence model star.

\begin{figure}
\begin{center}
\includegraphics[clip,width=0.9\columnwidth]{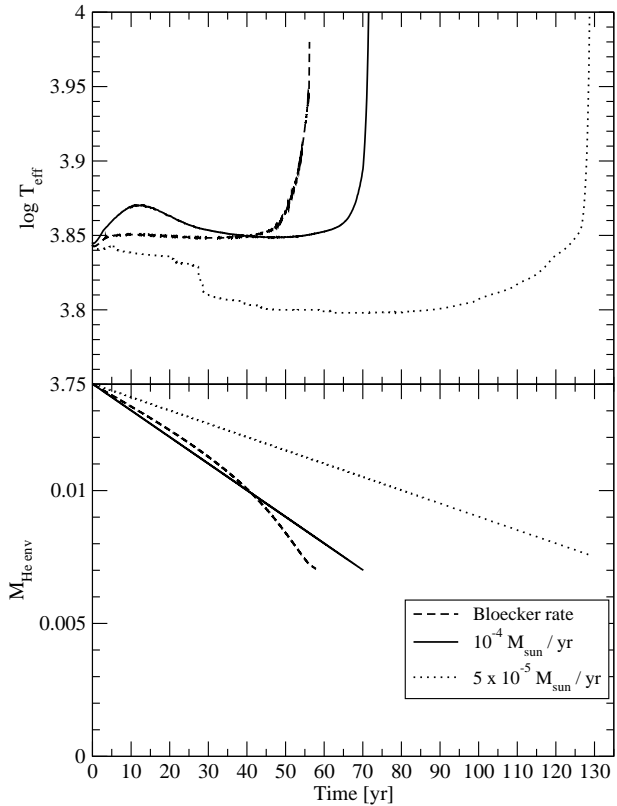}
\caption{The  effective temperature and  mass of  the He  envelope (in
         solar masses)  of the $0.741\, M_{\sun}$ model  sequence as a
         function  of  time   during  the  Sakurai  stage.   Different
         mass-loss  rates  as  indicated   in  the  legend  have  been
         assumed.}
\label{perfil}
\end{center}
\end{figure}

In Table 1 we show, for each stellar mass, the mass of the He envelope
of  our models  left shortly  after the  occurrence of  the born-again
episode --- that  is, by the time the Sakurai's  stage is reached.  We
define the He  envelope as the region of the  star where the abundance
mass fraction of He is larger  than 0.01.  We also list the luminosity
achieved  shortly  after  the  Sakurai  stage.  Note  that  for  these
sequences,  which cover  most of  the  observed mass  range of  PG1159
stars, the expected range of He envelope masses spans almost one order
of magnitude.  {\sl These values are upper limits because no mass loss
was introduced  in our evolutionary sequences after  the occurrence of
the VLTP}. Note as well that the most massive models are characterized
by He envelopes  whose masses are substantially smaller  than those of
the less massive  models.  In particular, for the  sequence with $0.87
\, M_{\sun}$ --- representative  of the spectroscopic mass of H1504+65
(Werner \& Herwig 2006; Miller Bertolami \& Althaus 2006) --- the mass
of the  He envelope  left amounts to  only $\sim 5.8  \times 10^{-3}\,
M_{\sun}$.  Finally, it  is also worth noting the  very large increase
of the maximum luminosities with increasing mass.

\begin{figure}
\begin{center}
\includegraphics[clip,width=0.9\columnwidth]{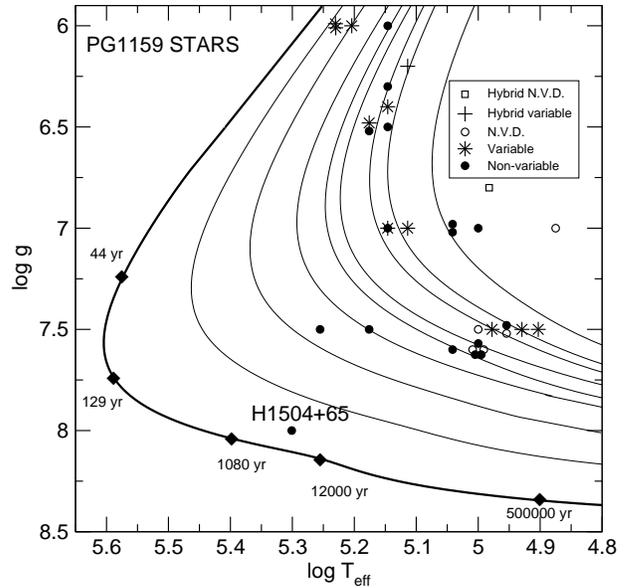}
\caption{Surface   gravity  versus   effective  temperature   for  the
         evolutionary sequences of  Miller Bertolami \& Althaus (2006)
         with  different stellar  masses ---  thin lines (0.741, 0.664,
         0.609, 0.589, 0.565, 0.542, 0.53, and $0.51\, M_{\sun}$, from left
         to right) ---  and the
         $0.87\, M_{\sun}$  model sequence with  thin He-rich envelope
         --- thick  line.  For  the latter,  post-born-again  ages are
         labelled along  the curve. The mass--loss rates  of Lawlor \&
         MacDonald (2006) have been adopted during the [WCL] stage.  The
         available observational  data for selected PG  1159 stars are
         also included. Spectroscopic  uncertainties are typically 0.5
         dex for the gravity and between 10 and 15\% for the effective
         temperature.}
\label{evolu}
\end{center}
\end{figure}

\section{A post-born-again origin for H1504+65?}

The fact  that the  most massive models  are characterized by  a small
He-rich  envelope is  particularly attractive  in view  of  the recent
evidence presented by  Van Hoof et al.  (2007)  about strong  mass losses
in Sakurai's  object,  which  has a  mass  of about  $0.60\,
M_{\sun}$.  Note  that if  the total mass  lost during  the Sakurai
stage is as high as $0.005\,M_{\sun}$, which is not discarded by
new radio and  optical observations of the Sakurai's  object (Van Hoof
et  al.  2007),  then  post-born-again remnants  as  massive as  about
$0.9\,  M_{\sun}$ are,  in principle,  expected to  lose  their entire
He-rich  envelope,  exposing   the  carbon-oxygen  core  during  their
subsequent  evolution to  the PG1159  regime (see  Table 1).    To
explore  this  possibility, we  have  computed  the  evolution of  the
$0.741$ and $ 0.87 \, M_{\sun}$ sequences from the born-again stage to
the  PG1159  phase by  assuming  stationary  mass  losses during  the
Sakurai and [WCL] stages.  For the Sakurai phase, we have considered the
mass-loss rate of Bl{\" o}cker (1995) and two constant mass-loss rates
of  $\dot  M=   5  \times  10^{-5}\,M_{\sun}/{\rm  yr}  $   and  $  M=
10^{-4}\,M_{\sun}/{\rm yr}$.  Strong  mass losses during the [WCL]
stage  have also been  reported by  Koesterke (2001).   During this
phase mass loss strongly depends on the  surface luminosity of the
remnant.  To  account for this we  have adopted the  mass-loss rate of
Lawlor  \& MacDonald  (2006).   In  particular, for  a  [WCL] star  with
$\log(L/L_{\sun})\ga     4.5$,     mass-loss     rates    of     $\sim
10^{-5}\,M_{\sun}/{\rm yr}$ are expected.

Our calculations show that,  irrespective of the mass-loss rate during
the  Sakurai stage,  the mass  of the  final He  envelope is  the same
independent  of the  mass-loss rate  prescription. Indeed,  only about
half the total mass of the  He-rich envelope can be eroded during this
stage.  This is  illustrated  in  Fig. \ref{perfil}  for  the case  of
$0.741\,  M_{\sun}$  sequence.   During  this stage  and  despite  the
distinct   evolutionary  timescales   resulting  from   the  different
mass-loss  prescriptions, the final  mass of  the He-rich  envelope is
only  reduced  from 0.014  to  about  $0.007\,  M_{\sun}$. Below a  critical  
$T_{\rm eff}$  mass  loss is  no  longer efficient in removing  the He 
envelope. In particular, note that
because of  the larger mass loss  provided by the  Bl{\" o}cker (1995)
formulation, departure  from the Sakurai stage occurs  earlier in this
case.  We  also found  that as  soon as the  envelope mass  is reduced
below  the critical value ---  of about $0.007\,  M_{\sun}$ 
in the case of our $0.741\,  M_{\sun}$  sequence  --- 
the subsequent evolution through  the [WCL]
stage proceeds  extremely fast  --- in a  few decades for  the $0.87\,
M_{\sun}$ sequence,  see Fig. \ref{evolu}.  As a result of  this rapid
evolution, the  mass lost from  the surviving He-rich  envelope during
the  [WCL]  stage   amounts  to  only  a  few   $  10^{-4}\,  M_{\sun}$,
independently  of the  exact amount  of mass  lost during  the Sakurai
stage.  This is much smaller than the final He-rich envelope mass with
which  the star  departs  from  its Sakurai  stage.  We conclude  that
stationary winds during  the post born again evolution  fail to remove
the  entire  He-rich  envelope  and  cannot explain  the  observed  He
deficiency of H1504+65. In  other words, the residual He-rich envelope
could  only be eroded  during the  [WCL] stage  as long  as most  of the
envelope is  lost during  the Sakurai stage,  which, according  to our
calculations,    is   not   possible    if   stationary    winds   are
assumed. Conversely, if the entire  He-rich envelope is to be removed,
 non-stationary mass loss episodes during the Sakurai stage are needed.

\section{Conclusions and discussion}

The evidence  presented in this investigation suggests  that  mass losses
during  the  Sakurai   stage  could  lead  to  PG1159  stars
characterized  by  He   envelopes  substantially  thinner  than  those
traditionally  accepted by  the standard  theory of  stellar evolution
(Herwig 1999, Miller Bertolami \& Althaus 2006). Motivated by the less
massive  He envelope  left by  prior evolution  of  massive born-again
stars, we  have explored  the possibility that  these stars  may loose
almost  all of  their residual  envelopes  as a  result of  stationary
mass--loss.  To this end,  we have computed the full evolution
of massive  post-born-again stars by considering  a variable mass-loss
history as well as constant  mass-loss rates during the Sakurai stage.
We find that stationary  winds during the post-born-again evolution do
not completely  remove the He-rich envelope and,  thus, cannot explain
the  absence of  He in  H1504+65.  Stationary  mass-losses  during the
Sakurai  stage only  remove at  most a  half of  the  He-rich envelope
surviving the  violent H  burning during the  born-again phase,
and the following evolution during  the [WCL] stage proceeds too fast to
erode by [WCL] winds the remaining envelope.  This result is independent
of the detailed treatment of mass loss during the Sakurai stage. Thus,
the  origin of  the recently  discovered white  dwarf  population with
carbon-dominated atmospheres cannot be traced back to VLTP progenitors
that have  undergone {\sl stationary} mass-loss during their
Sakurai stage. Conversely, for  H1504+65 to represent the evolutionary
link  between  born-again  stars  and white  dwarfs  with  carbon-rich
atmospheres,  most  of  the  He-rich  envelope should  be  ejected  by
non-stationary winds  during the  Sakurai stage.  This  possibility is
not entirely  discarded since ejection  of the whole  He-rich envelope
should occur from the layers at  which the thermal energy input of the
He  flash takes place  and above  which our  evolutionary calculations
show  that the  envelope rapidly  expands. The  feasibility  that {\sl
most}  of the  envelope could  be ejected  in more  massive  models is
strengthened by the fact that  a massive remnant is characterized by a
markedly larger  surface luminosity  (by a factor  of 4) and  a larger
input  energy  from the  He  flash,  as  compared with  the  Sakurai's
object. In  closing, we  note  that the  possibility 
that traces  of He
remain in H1504+65  cannot be totally discarded.  If  some He survives
the mass-loss events,  diffusion will lead to a  He-rich white dwarf
and the carbon-rich atmosphere should emerge as a result of convective
dredge-up   at   smaller   effective  temperatures.   Exploring   this
possibility  requires a  more  elaborated treatment  of the  evolution
during the Sakurai stage and is beyond the scope of the present work.

\begin{acknowledgements}
We acknowledge useful discussions  with M. M.  Miller Bertolami.  Part
of this work was supported by the MEC grant AYA05-08013-C03-01, by the
European Union  FEDER funds, by the  AGAUR, by AGENCIA through
the Programa de Modernizaci\'on Tecnol\'ogica BID 1728/OC-AR,
and by PIP  6521 grant from CONICET.
\end{acknowledgements}


\begin{thebibliography}{}
\bibitem{Aea08}  Althaus, L. G.,  C\'orsico, A. H.,  Miller Bertolami,
                 M.  M.,  Garc\'{\i}a--Berro, E.,  \&  Kepler, S.  O.
                 2008, ApJL, 677, L35
\bibitem{Aea05}  Althaus,  L. G.,  Serenelli,  A.  M.,  Panei, J.  A.,
                 C\'orsico,    A.  H.,   Garc\'{\i}a--Berro,   E.,  \&
                 Sc\'occola,  C. G. 2005, A\&A, 435, 631
\bibitem{B01}    Bl{\" o}cker, T. 1995, A\&A, 297, 727
\bibitem{B01}    Bl{\" o}cker, T. 2001, Ap\&SS, 275, 1
\bibitem{Ceaip}  Costa, J. E. S.,  \&  Kepler,  S. O.  2008, A\&A, 489,
                 1225
\bibitem{DH98}   Dreizler, S., \& Heber, U. 1998, {A\&A}, 334, 618
\bibitem{D07}    Dufour, P., Liebert, J., Fontaine, G., \& Behara, N.
                 2007, Nature, 450, 522
\bibitem{D08}    Dufour, P., Fontaine, G., Liebert, J., Schmidt, G.D., 
                 \&    Behara,    N. 2008    \apj, 683, 978 
\bibitem{GB97}   Garc\'\i a--Berro, E., Ritossa, C., \&  Iben, I. 1997,
                 ApJ, 485, 765
\bibitem{Gea04}  Guerrero,  J., Garc\'\i a--Berro, E.,  \& Isern,  J.
                 A\&A, 2004, 257
\bibitem{h99}    Herwig, F., Bl\"ocker, T., Langer N., \& Driebe, T.  
                 1999, A\&A, 349, L5
\bibitem{IEA83}  Iben,  I., Kaler, J. B.,  Truran,  J. W., \& Renzini, 
                 A. 1983, ApJ, 264, 605
\bibitem{K01}    Koesterke, L. 2001, Ap\&SS, 275, 41
\bibitem{LM}     Lawlor T. M., \& MacDonald J. 2006, MNRAS, 371, 236
\bibitem{Mea06}  Miller Bertolami,  M. M.,  \& Althaus,  L.  G. 2006,
                 A\&A, 454, 845
\bibitem{R06}    Rauch, T. 2006, in  {\sl ``Planetary Nebulae in our Galaxy
                 and Beyond''}, IAU Symposium, 234, 131
                 in press, {\tt arXiv: 0711.4565v1} 
\bibitem{Rea96}  Ritossa, C.,  Garc\'{\i}a--Berro,   E., \& Iben, I. 1996,
                 ApJ, 460, 489
\bibitem{S79}    Sch\"onberner, D. 1979, A\&A, 79, 108
\bibitem{S92}    Sch\"onberner, D., \& Bl\"ocker, T. 1992, in  
                 {\sl ``The atmospheres of Early--Type Stars''}, Lectures
                 Notes in Physics (Springer), 401, 305
\bibitem{VH07}   Van  Hoof, P. A. M.,  et al. 2007, A\&A, 471, L9
\bibitem{W01}    Werner, K. 2001, Ap\&SS,  275, 27
\bibitem{W06}    Werner, K., \& Herwig, F. 2006, PASP, 118, 183
\bibitem{W04}    Werner, K., Rauch, T., Barstow, M. A., \& Kruk, J. W. 2004, 
                 A\&A, 421, 1169
\bibitem{W08}    Werner, K., Rauch, T., Reiff, E., \& Kruk, J. W. 2008, 
                 in {\sl ``Hydrogen-deficient stars''}, ASP Conf. Ser., 391, 
                 109
\end{thebibliography}
\end{document}